\documentclass[twoside,leqno,twocolumn]{article}

\usepackage{ltexpprt}
\usepackage{booktabs} 
\usepackage{algorithm}
\usepackage{algpseudocode}
\usepackage{graphicx}
\usepackage{caption}
\usepackage{subcaption}
\usepackage{color}
\usepackage{hyperref}
\usepackage{amsmath}
\usepackage{tabularx}
\usepackage{amsfonts}
\usepackage{xspace}
\usepackage{multirow}
\usepackage{cite}
\usepackage{footnote}
\usepackage[table,xcdraw]{xcolor}
\usepackage{mdwlist}
\makesavenoteenv{tabular}

\newcommand{\method}{UniWalk\xspace}
\newcommand{\fullname}{Explainable and Accurate Recommendation for Rating and Network Data}
\newtheorem{mydef}{Definition}
\definecolor{orange}{rgb}{1, 0.5, 0}
\definecolor{purple}{rgb}{0.64, 0.25, 0.84}
\floatname{algorithm}{Algorithm}

\newtheorem{problem}{Problem}

\begin{document}

\title{\Large \method: \fullname}

\author{
	Haekyu Park
	\thanks{Seoul National University, Seoul, Republic of Korea.
				\{hkpark627, jeon185, kjh900809, elaborate, ukang\}@snu.ac.kr} \\
	\and
	Hyunsik Jeon\footnotemark[1]
	\and
	Junghwan Kim\footnotemark[1]
	\and
	Beunguk Ahn\footnotemark[1]
	\and
	U Kang\footnotemark[1]}
\date{}
\maketitle


\setlength{\floatsep}{0.03cm}
\setlength{\textfloatsep}{0.03cm}
\setlength{\intextsep}{0.03cm}
\setlength{\dblfloatsep}{0.03cm}
\setlength{\dbltextfloatsep}{0.03cm}
\setlength{\abovecaptionskip}{0.03cm}
\setlength{\belowcaptionskip}{0.03cm}
\setlength{\abovedisplayskip}{0.03cm}
\setlength{\belowdisplayskip}{0.03cm}

\begin{abstract} 
How can we leverage social network data and observed ratings to correctly recommend proper items and provide a persuasive explanation for the recommendations?
Many online services provide social networks among users, and
it is crucial to utilize social information since recommendation by a friend is more likely to grab attention than the one from a random user.
Also, explaining why items are recommended is very important in encouraging the users' actions such as actual purchases.
Exploiting both ratings and social graph for recommendation, however, is not trivial because of the heterogeneity of the data.

In this paper, we propose \method, an explainable and accurate recommender system that exploits both social network and rating data.
\method combines both data into a unified graph,
learns latent features of users and items,
and recommends items to each user through the features.
Importantly, it explains why items are recommended together with the recommendation results.
Extensive experiments show that \method provides the best explainability
and achieves the state-of-the-art accuracy. 
\end{abstract}

\section{Introduction}
Given rating and social network data, how can we recommend appropriate items convincingly?
Recommending proper items is a crucial task benefiting both suppliers and consumers.
Recommendation increases sales for suppliers by capturing latent demands and facilitates searches over massive items for consumers.
Providing reasons behind a recommendation is much more important since it enriches user experiences and draws out user participation,
which eventually enhances long-term performance of the recommender system.

\begin{figure*}[htbp]
	\centering
	\begin{subfigure}[b]{0.30\textwidth}
		\centering
		\includegraphics[width=1\textwidth]{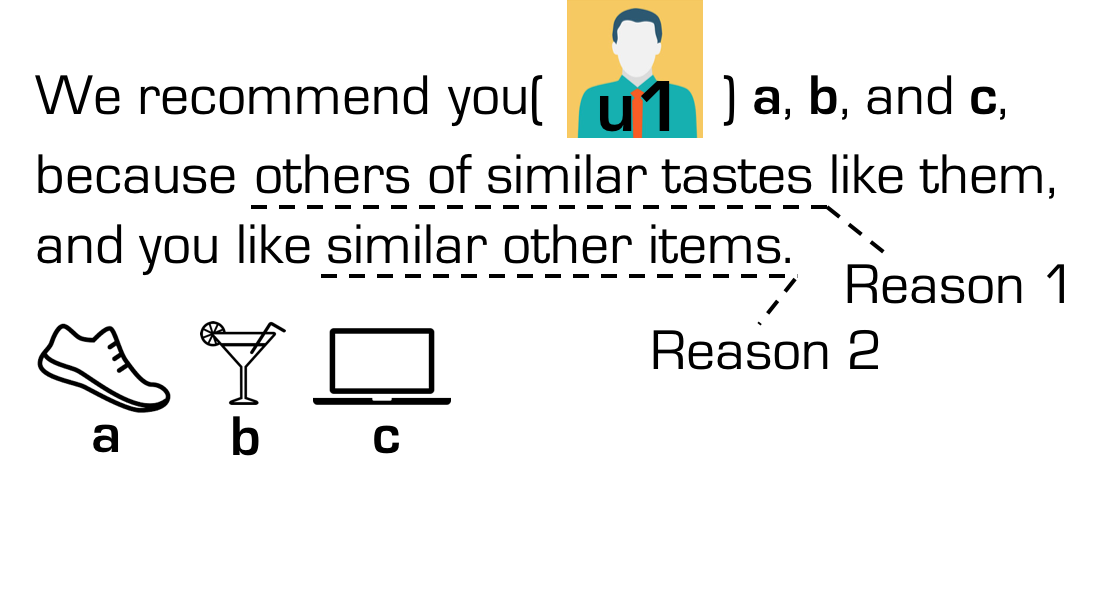}
		\caption{
			We recommend items ($a$, $b$, and $c$) to a target user ($u1$) with two reasons (Figures \ref{fig:reason1} and \ref{fig:reason2}).
		}
		\label{fig:explain}
	\end{subfigure}
	~
	\begin{subfigure}[b]{0.35\textwidth}
		\centering
		\includegraphics[width=0.8\textwidth]{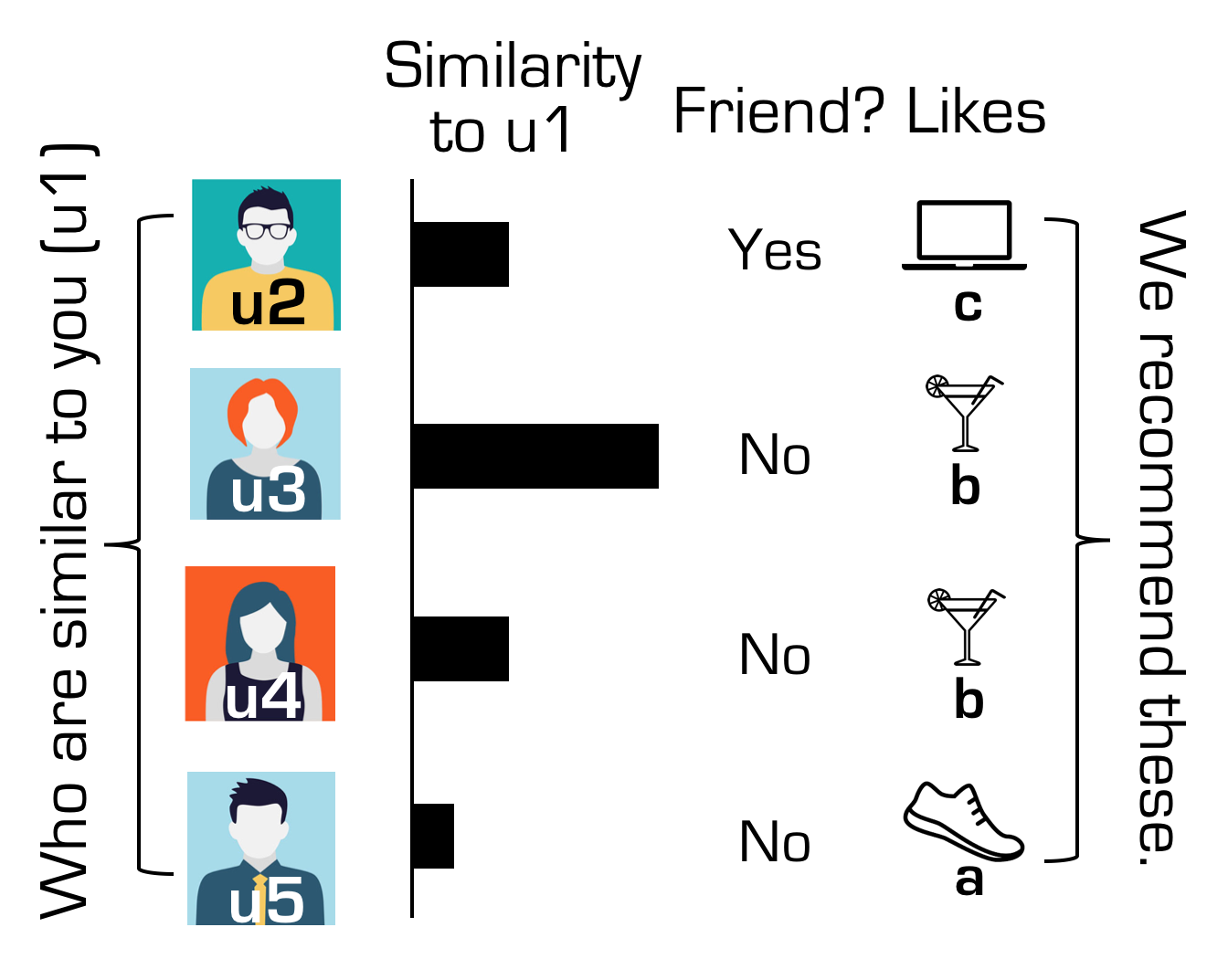}
		\vspace{-0.7em}
		\caption{
			Reason1: Explain with similar users. The recommended items are preferred by other users ($u2$, $u3$, $u4$, and $u5$) who are similar to the target user.
		}
		\label{fig:reason1}
	\end{subfigure}
	~
	\begin{subfigure}[b]{0.30\textwidth}
		\centering
		\includegraphics[width=0.7\textwidth]{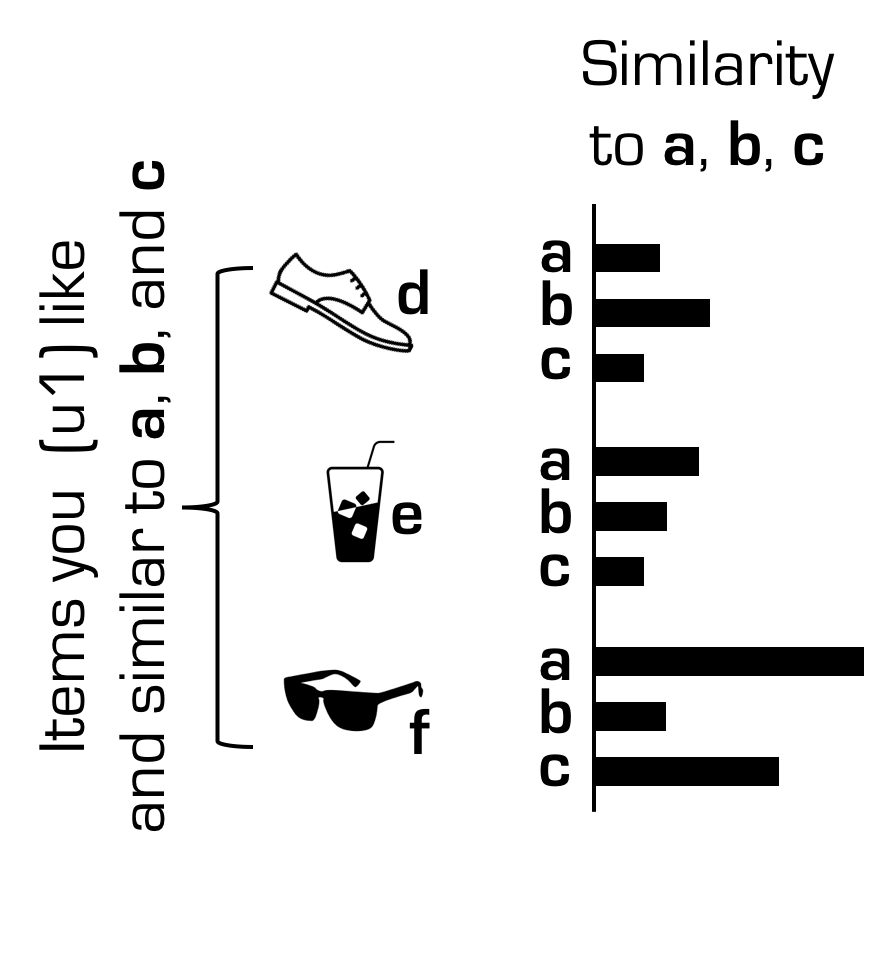}
		\caption{
			Reason2: Explain with similar items. The target user likes other similar items ($d$, $e$, and $i$).
		}
		\label{fig:reason2}
	\end{subfigure}
	\caption{
		\method explains its recommendation using two reasons.
		More details are described in Section \ref{sec:exp:explain}.
	}
	\label{fig:Explainability}
\end{figure*} 	

Most recommender systems are based on rating data only.
However, rating data are sparse and expensive to gather.
Additional information is required to overcome the sparsity and enhance the recommendation results.
One of the most prevalent additional information accompanying e-commerce services is social network among users.
Social influence is powerful in recommendation, since people trust recommendation through acquaintance more than that from an anonymous one.
Social information also provides useful information on tastes of users according to homophily property \cite{mcpherson2001birds}: neighbors in social networks are likely to share similar preferences.
Consequently, it is essential to exploit social networks as well as rating data for explainable and accurate recommendations.

\begin{table*}[t]
	\vspace{0.3em}
	\centering
	\caption{Comparison of \method and other methods. \method provides the best explainability and accuracy, and it is the only method that utilizes all types of links. Meta-explanation is sub-explanation for the main explanation.} 
	\label{table:comp}
	\renewcommand{\tabcolsep}{3mm}
	\small
	\begin{tabular}{c  c  c  c  c  c  c}
		\toprule
		& \multicolumn{1}{c}{Methods $\rightarrow$}           & \multicolumn{1}{|c}{\textbf{UniWalk (Proposed)}}         & \multicolumn{1}{c}{UCF} & \multicolumn{1}{c}{ICF}  & \multicolumn{1}{c}{TrustSVD}  & MF  \\ \midrule
		\multirow{5}{*}{Explainability}                           & \multicolumn{1}{|c|}{Can explain}
		& \textbf{Yes}  & Yes & Yes & No & No\\
		& \multicolumn{1}{|c|}{Meta-explanation}
		& \textbf{Yes}  & No & No  & No  & No   \\
		& \multicolumn{1}{|c|}{Using similar users}
		& \textbf{Yes}  & Yes & No & No & No   \\
		& \multicolumn{1}{|c|}{Using friends}
		& \textbf{Yes} & No & No & No & No   \\
		& \multicolumn{1}{|c|}{Using similar items}
		& \textbf{Yes} & No & Yes & No & No  \\ \midrule
		\multirow{2}{*}{Error} & \multicolumn{1}{|c|}{RMSE} & \multicolumn{1}{c}{\textbf{Lowest (most accurate)}} & \multicolumn{1}{c}{Low}  & \multicolumn{1}{c}{Low} & \multicolumn{1}{c}{High} & High \\
		& \multicolumn{1}{|c|}{MAE}        & \multicolumn{1}{c}{\textbf{Lowest (most accurate)}} & \multicolumn{1}{c}{Low}      & \multicolumn{1}{c}{Low} & \multicolumn{1}{c}{High} & High \\ \midrule
		\multicolumn{1}{l}{\multirow{3}{*}{Utilization of links}} & \multicolumn{1}{|c|}{Ratings}
		& \textbf{Yes} & Yes & Yes & Yes & Yes  \\
		& \multicolumn{1}{|c|}{Social links}
		& \textbf{Yes}  & No & No & Yes  & No   \\
		& \multicolumn{1}{|c|}{Multi-step links}
		& \textbf{Yes} & No & No & No & No  \\
		\bottomrule
	\end{tabular}
	\vspace{0.3em}
\end{table*}

Leveraging both rating and social network data in recommendation and explanation, however, is challenging.
Many multi-step relations such as friends of friends or friends' favorite items provide useful information on users' tastes,
but there is no straightforward mechanism to translate the multi-step links into rating values.
In addition, latent features represented by many collaborative filtering methods are no more than series of numbers to human.
Existing methods suffer from those difficulties.
For example, recommendation methods such as TrustSVD \cite{guo2015trustsvd}, ASMF and ARMF \cite{li2015point} use social networks partially.
TrustSVD utilizes only direct friendships, and
ASMF and ARMF use social graphs only to pick out potential items in a pre-processing stage.
Also, those methods do not offer any explanations.

In this paper, we propose \method, an explainable and accurate recommendation model that utilizes both ratings and graph data.
\method generates a unified graph describing both preference and user similarity.
Then, it samples not only quantitative links representing rating data but also qualitative links indicating entities' similarity or dissimilarity from the unified graph with random walks.
Next, \method learns entities' latent features with a newly devised objective function consisting of a supervised term for the quantitative links
and unsupervised terms for the qualitative links.
\method then recommends items of the highest predicted ratings and provides reasons of the recommendation.
\method provides the best explainability and accuracy among competitors, as shown in Table~\ref{table:comp}.

The main contributions of this paper are as follows.

\begin{itemize*}
	\item \textbf{Method.}
		We propose \method, a novel method that exploits ratings and social graph for recommendations.
		\method leverages not only quantitative rating information but also qualitative similarity and dissimilarity information for better understanding of users and items.
	
	\item \textbf{Explainability.}
		\method explains why items are recommended to a target user (Figure \ref{fig:Explainability}).
		\method recommends items when similar users like the same items 
		or the target user prefers other similar items. 
		\method also provides meta-explanations for better persuasiveness: why the other users and items are determined to be similar to the target user and the recommended item, respectively. 
	
	\item \textbf{Accuracy.}
		\method achieves the state-of-the-art accuracy for recommendation with ratings and a social network in terms of RMSE and MAE (Section \ref{sec:accuracy}).

\end{itemize*}

The rest of this paper is developed as follows.
We first explain backgrounds of \method including matrix factorization and network embedding in Section \ref{sec:prelim}.
Then we propose our method \method in Section \ref{sec:proposed}.
After presenting experimental results in Section \ref{sec:experiment},
we list related works in Section \ref{sec:related}
and conclude this paper in Section \ref{sec:conclusion}.
The codes and datasets are available at \url{http://datalab.snu.ac.kr/uniwalk}.

\section{Preliminary}
\label{sec:prelim}
\label{sec:notation}
In this section, we describe preliminaries on matrix factorization and network embedding.

\vspace{-1em}
\subsection{Matrix Factorization with Bias Factors.}
\label{sec:mf}
Matrix factorization (MF) predicts unobserved ratings given observed ratings.
MF predicts a rating of an item $i$ given by a user $u$ as
$\hat{r}_{ui} = \mu + b_u + b_i + \mathbf{x}_u^T \mathbf{y}_i$,
where $\mu$ is global average rating, $b_u$ is $u$'s bias, $b_i$ is $i$'s bias,
$\mathbf{x}_u$ is $u$'s vector, and $\mathbf{y}_i$ is $i$'s vector.
Biases indicate the average tendency of rating scores given by users or given to items.
The bias terms are known to boost the performance of prediction accuracy compared to that by inner product term only.

The objective function is defined in Equation \eqref{eq:L}, where $r_{ui}$ is an observed rating and $K$ is a set of (user, item) pairs
for which ratings are observed.
The term $\lambda(b_u^2 + b_i^2 + ||\mathbf{x}_u||^2 + ||\mathbf{y}_i||^2)$ prevents overfitting by regularizing the magnitude of parameters,
where the degree of regularization is controlled by the hyperparameter $\lambda$.
The objective function is minimized by gradient descent.
\vspace{-0.6em}
\begin{align}
\label{eq:L}
\small
L = \frac{1}{2}\sum_{(u,i) \in K}
&(r_{ui} - \mu - b_u - b_i - \mathbf{x}_u^T \mathbf{y}_i)^2\\ \notag
&+ \lambda(b_u^2 + b_i^2 + ||\mathbf{x}_u||^2 + ||\mathbf{y}_i||^2)
\end{align}

\subsection{Network embedding.}
\label{sec:ge}
Network embedding represents each node as a vector that encodes structural information on the network.
The task can be formalized as follows: given a graph $G = (V, E)$, find a function $f: V \rightarrow \mathbb{R}^d$
mapping each node $v \in V$ to a $d$-dimensional vector $f(v)$.

DeepWalk \cite{perozzi2014deepwalk} is one of the most representative network embedding methods.
It exploits Skipgram with negative sampling (SGNS) to learn node vectors.
DeepWalk simulates multiple truncated random walks to apply SGNS on network data.
For each node in the network, the model generates random walks of fixed length.
Each step of random walk selects next node according to transition probabilities defined by weights on edges.
The resulting walk is a sequence of nodes in which more strongly connected node pairs appear in close distance more frequently.
Then, Skipgram defines `neighborhood' where a pair of nodes co-occurs in the simulated random walk.
Neighborhood node pair is defined by a window sliding on the node sequence.
The center node in the window is defined as a $target$, and other nodes
in the window are $neighbors$ of the $target$.
Pairs of co-occurring $targets$ and $neighbors$ are stored in a set $\mathcal{D}$.
Co-occurrence statistic represents proximity between nodes in the graph, since nodes closer in the graph co-occur more frequently in the random walk.

If the model optimizes over only proximate pairs, the embedding would converge to a single point.
SGNS model applies negative sampling to prevent this convergence of embedding process
by adding distant pairs into the process.
Negative sampling technique randomly selects pairs of nodes and pushes embeddings for those pairs apart from each other.
For each proximate node pair $(v, w) \in \mathcal{D}$,
$n$ randomly selected nodes $w_1, ..., w_n$ form $n$ distant node pairs $(v, w_1), ..., (v, w_n)$.

\vspace{-1em}
\section{Proposed Method}
\label{sec:proposed}
In this section, we describe \method, our proposed method for rating prediction
and explanation on its logic behind recommendations.
We solve the following problems.
\vspace{-0.3em}
\begin{problem} \textsc{(Recommendation with Ratings and a Graph)}
	Given an explicit feedback rating matrix $\mathbf{R} \in \mathbb{R}^{|U| \times |I|}$ and a social network $G = (V, E)$,
	predict a rating of an item $i$ given by a user $u$ which is denoted as $\hat{r}_{ui}$.
	Ideally, $\hat{r}_{ui}$ is close to the corresponding observed rating $r_{ui} = \mathbf{R}_{ui}$.
	\label{prob:recommend}
\end{problem}
\vspace{-0.4em}
\begin{problem} \textsc{(Explanation of Recommendation)}
		For a user $u$ and predicted rating matrix $\hat{\mathbf{R}} \in \mathbb{R}^{|U| \times |I|}$,
		recommend items of the highest predicted ratings and explain why they are recommended to $u$.
		\label{prob:explain}
\end{problem}

In the remaining part of this section, we first state challenges for explainable and accurate recommendation
and our main ideas to overcome the challenges in Section \ref{sec:challenges_and_solutions}.
Then, we describe our proposed method \method in detail to solve Problem \ref{prob:recommend} in Section \ref{sec:accurate_method} and to solve Problem \ref{prob:explain} in Section \ref{sec:explain}.

\vspace{-1em}
\subsection{Challenges and Solutions.}
\label{sec:challenges_and_solutions}
For explainable and accurate recommendation, we take both quantitative and qualitative relationships of users and items into account.
Quantitative links represent rating scores, and qualitative links indicate similarity or dissimilarity among users and items.
They are stated as follows.
\vspace{-0.6em}
\begin{mydef} Link types for our learning process.
	\vspace{-0.6em}
	\label{def:links}
	\begin{itemize*}
		\item Quantitative or score links are edges between users and items that represent observed ratings.
		\item Qualitative links are relationships among users and items to represent similarity or dissimilarity among them. There are two types of qualitative links.
		\begin{itemize*}
			\vspace{-0.5em}
			\item Similarity links are relationships among similar entities.
			They include friend links and multi-step links between similar entities such as items rated similarly by common users.
			\item Dissimilarity links are multi-step links among dissimilar entities such as users and their friends' unfavorable items.
		\end{itemize*}
	\end{itemize*}
\end{mydef}
\vspace{-0.6em}

Modeling qualitative links is beneficial in two ways.
First, the similarity links are used in explaining reasons of recommendation. We recommend items since ``similar" other users like them, or the recommended items are ``similar" to other observed preferred items.
Second, the qualitative links lead to more accurate recommendation.
The qualitative links contain additional information on users' tastes and items' properties and mitigate data-sparsity problem in rating and social data.

However, there are three key difficulties in using those qualitative links:
(1) drawing out proper qualitative links spanning both ratings and a social network,
(2) learning from the qualitative links in accordance with the score links,
and (3) using qualitative links systemically in explanation.
Most real-world graphs show the small-world phenomenon, meaning that the diameters are small,
and thus random entities can be paired even with a small number of steps;
sampling appropriate pairs of similar and dissimilar entities is essential for the learning process.
In addition, qualitative links only imply possible similarities or dissimilarities in properties among users and items, while a quantitative link directly gives numerical values, which we predict.

We provide solutions to those challenges.
\method uses efficient random walk sampling to pick similar or dissimilar entities to tackle the first challenge.
Proximate entity pairs in a simulated random walk are our desired qualitative connected pairs which potentially contain informative properties of entities.
Random walk sampling process is efficient, because each sampled entity in the sequence of random walk forms multiple connections by being paired with multiple nearby entities.
The second challenge is solved by our objective function that covers quantitative and qualitative links.
The objective function combines supervised learning for quantitative links and unsupervised learning for qualitative ones in the same framework.
We solve the last challenge by measuring similarities of the qualitative links and presenting the numerical values in explanation.

\vspace{-1em}
\subsection{\method for Rating Prediction.}
\label{sec:accurate_method}
We present details of the process of \method to solve Problem \ref{prob:recommend}.
\method comprises three steps.
First, it generates a unified graph to leverage rich relations among users and items from the heterogeneous input data.
Next, \method applies network embedding on the unified graph to learn bias and vector encoding preference or properties of users and items.
Lastly, it calculates predicted ratings $\hat{r}_{ui}$ for each item $i$ and user $u$.

\paragraph{Step 1: Build a unified graph.}
\label{sec:stepgb}
In this step, we generate a weighted and undirected graph $\tilde{G}$ that combines ratings and a social network whose weights of edges present similarity of entities.
Each rating is represented by a user-item edge in $\tilde{G}$.
Social links are described as user-user edges in $\tilde{G}$.
High ratings convey similarities of the users and items, and friends' similarities are determined by a hyperparameter $c$.
The process of generating $\tilde{G} = (\tilde{V}, \tilde{E})$ is as follows.
First, we convert rating matrix $\mathbf{R}$ into a graph structure.
For a user $u$ and an item $i$ where a rating $r_{ui}$ is observed,
we add $u$ and $i$ into $\tilde{V}$ and $(u, i)$ into $\tilde{E}$.
The added edge $(u, i)$ has a weight of $r_{ui}$.
Next, we add edges in a social network $G$ into $\tilde{G}$.
Weights of the added social edges are set to the hyperparameter $c$.
Users and items are termed as \textit{entities} in the unified graph.

\begin{mydef} Entity.\\
	Entities are users and items in the unified graph of ratings and a social network.
\end{mydef}

\paragraph{Step 2: Embed nodes of $\tilde{G}$.}
\label{sec:stepge}
\method applies network embedding on the unified graph $\tilde{G}$ to learn biases and latent vectors of entities.
There are three goals in our embedding method.
First, we aim to reduce differences between observed ratings and predicted ratings corresponding to score links.
Second, we increase inner products of vectors of two entities related by similarity links.
Third, we decrease inner products of vectors of dissimilar entities connected by dissimilarity links.

We devise a loss function $L$ consisting of terms \eqref{loss1} and \eqref{loss23}.
The terms implement the above three goals.
Term \eqref{loss1} is aimed to minimize the difference between an observed rating $r_{ui}$ and a predicted rating $\hat{r}_{ui}$ for a user $u$ and an item $i$.
The left term in Equation \eqref{loss23} drives latent vectors $\mathbf{z}_v$ and $\mathbf{z}_w$ of similar entities $v$ and $w$ to have a high inner product.
The right term in Equation \eqref{loss23} is aimed to make latent vectors $\mathbf{z}_v$ and $\mathbf{z}_w$ of dissimilar entities $v$ and $w$ have a negatively high inner product.
We call the term \eqref{loss1} as supervised term,
the left term in Equation \eqref{loss23} as positive unsupervised term, and the right one in Equation \eqref{loss23} as negative unsupervised term.
In the objective function $L$, a predicted rating $\hat{r}_{ui}$ is calculated as $\mu + b_u + b_i + \mathbf{z}_u^T\mathbf{z}_i$, where $\mu$ is the average of rating values,
$b_v$ and $\mathbf{z}_v$ are a bias factor and a latent vector of entity $v$, respectively.
$\lambda_b$ and $\lambda_z$ are regularization parameters for bias factors and latent embedding vectors, respectively.
$\alpha$ and $\beta$ indicate weight of the positive unsupervised term and the negative unsupervised term, respectively.
$\mathbf{b} \in \mathbb{R}^{|\tilde{V}|}$ is the bias vector of all entities, 
and $\mathbf{Z} \in \mathbb{R} ^ {|\tilde{V}|\times d}$ is the latent vector matrix of entities where $d$ is the dimension of entities' vector $\mathbf{z}_v$.
\vspace{-0.5em}
\begin{equation}
\small
L=
\overbrace {
	\sum_{(u, i) \in \mathcal{D}^R}
	\frac{1}{2} (r_{ui} - \hat{r}_{ui})^2
} ^ {\text{Supervised term}}
+ \overbrace{
	\frac{\lambda_b}{2} ||\mathbf{b}||^2
	+ \frac{\lambda_z}{2} ||\mathbf{Z}||^2_F
} ^ {\text{Regularization term}}
\label{loss1}
\end{equation}
\vspace{-1.5em}
\begin{equation}
\small
+\alpha \cdot \hspace{-7mm}
\overbrace{
	\sum_{(v, w) \in \mathcal{D}^+}
	-\mathbf{z}_v^T\mathbf{z}_w
} ^ {\text{Positive unsupervised term}}
+\hspace{3mm}\beta \cdot \hspace{-7mm}
\overbrace{
	\sum_{(v, w) \in \mathcal{D}^-}
	\mathbf{z}_v^T\mathbf{z}_w
} ^{\text{Negative unsupervised term}}
\label{loss23}
\end{equation}	

Step 2 mainly consists of two parts: sampling and optimization.
We sample node sequences from the unified graph in the sampling phase.
We extract entity pairs from the sequences and minimize $L$ in the optimization phase.

In the sampling phase, we generate node sequences with the following random walks on the unified graph.

\begin{figure}[t]
	\centering
	\begin{subfigure}[b]{0.5\textwidth}
		\centering
		\includegraphics[width=0.5\textwidth]{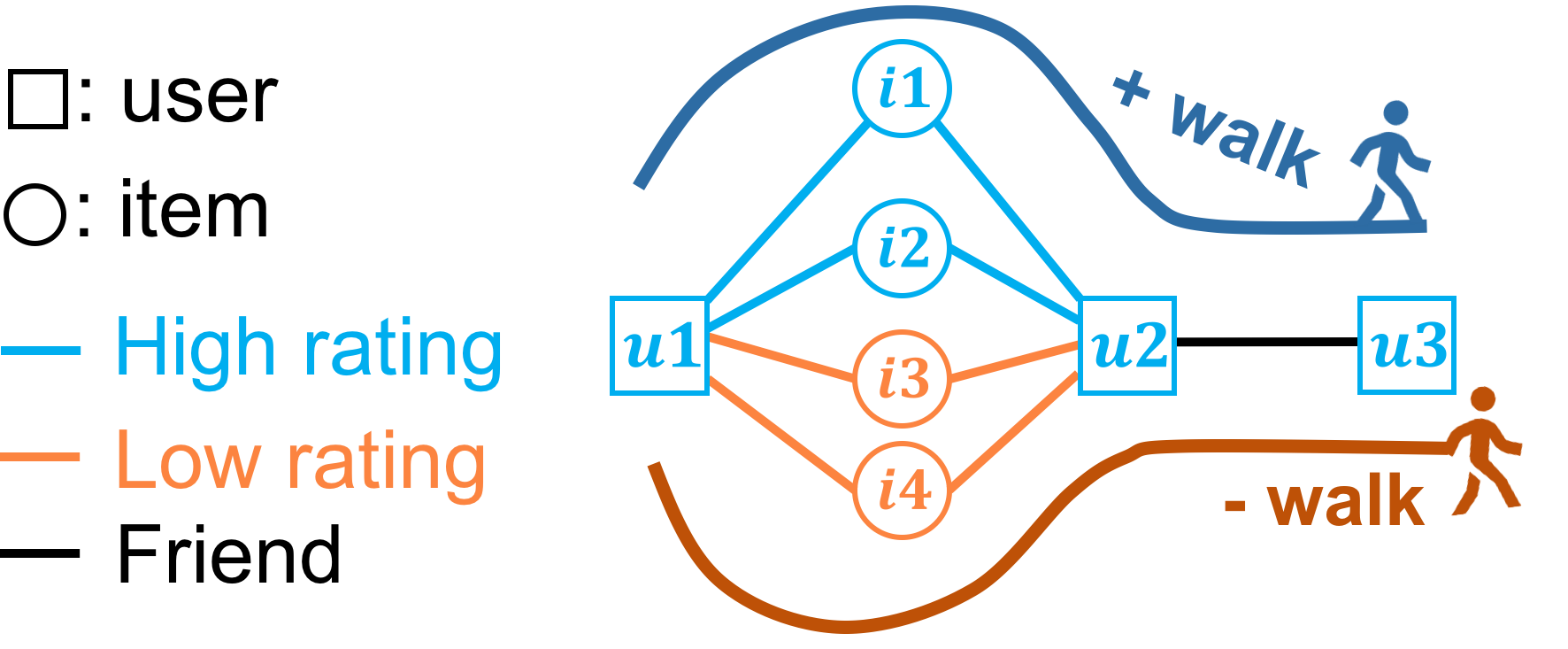}
		\vspace{-0.5em}
		\caption{
				Sampling phase to sample node sequences with random walks.
				The positive walk (+walk) is likely to walk on the blue high rating links and the black social link.
				The negative walk (-walk) is likely to walk on the orange low rating links and the black friend link.
				Unweighted walk is omitted, since it samples sequences randomly.
		}
		\label{fig:ex1}
	\end{subfigure}
	~
	\begin{subfigure}[b]{0.5\textwidth}
		\centering
		\includegraphics[width=0.57\textwidth]{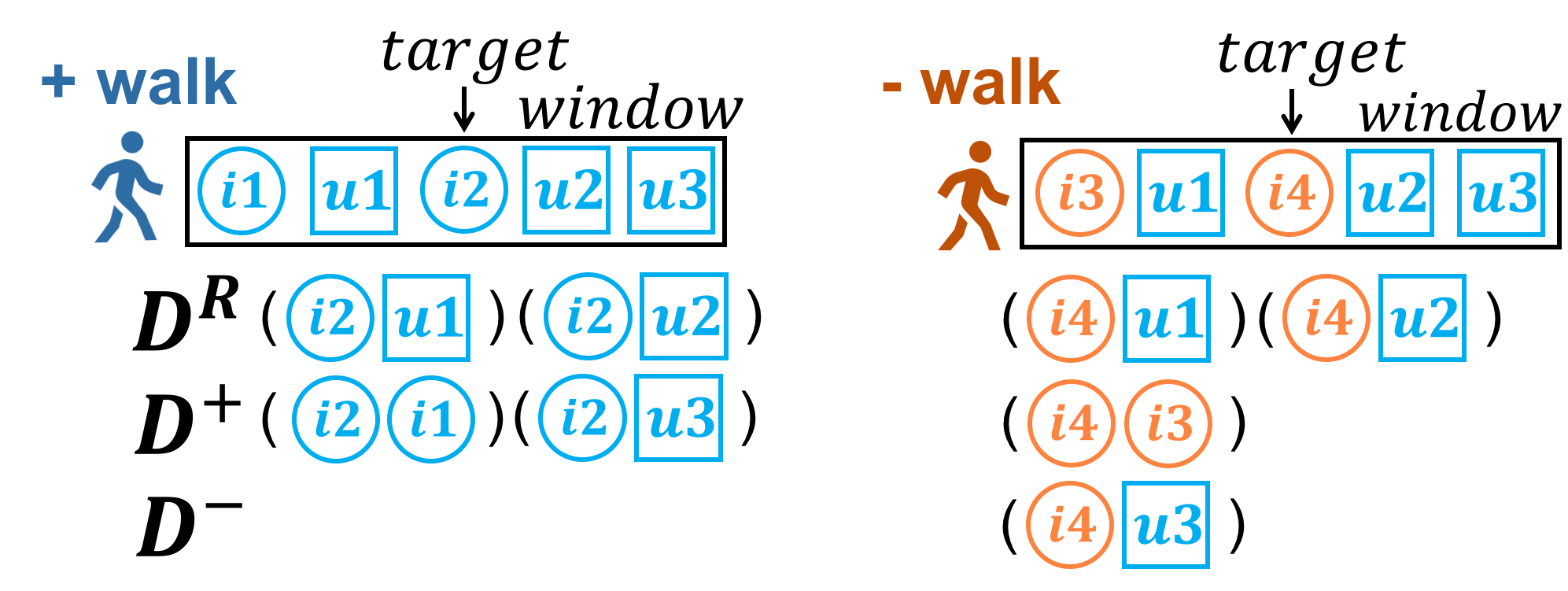}
		\vspace{-0.5em}
		\caption{
				Extracting node pairs in the optimization phase.
				$\mathcal{D}^R$, $\mathcal{D}^+$, and $\mathcal{D}^-$ are for entities linked by ratings, similar entities, and dissimilar entities, respectively.
		}
		\label{fig:ex2}
	\end{subfigure}
	\caption{
			An example of step 2 in \method.
			(a) shows the sampling phase,
			and (b) shows extracting pairs in the optimization phase.
	}
	\label{fig:example}
\end{figure}

\begin{mydef} Random walks for sampling links.
	\label{def:walks}
	\begin{itemize*}
		\vspace{-0.5em}
		\item Positive random walk is a random walk whose transition probabilities are proportional to weights on edges.
		\item Negative random walk is a random walk whose transition probabilities are proportional to (minR + maxR - weights on edges) for score links, and weights on edges for social links, where maxR and minR are the maximum and the minimum ratings, respectively.
		\item Unweighted random walk is a random walk whose transition probabilities are uniform in all edges.
	\end{itemize*}
\end{mydef}
The positive random walk generates sequences of similar entities with its transition rule, since high weight of an edge indicates high similarity between the nodes.
The negative random walk generates sequence of entities of two groups: users who dislike similar items and the items not preferred by the user group.
The groups are dissimilar to each other, and members in a group are similar.
The unweighted walk has the effect of sampling more substantial nodes, because higher degree nodes, which are active users or items, are sampled more.
Figure \ref{fig:ex1} shows an example of the sampling phase.

In the optimization phase, we extract node pairs from the sampled node sequences and learn features of the nodes.
We define sets of the node pairs as follows.
\begin{mydef} Entity pair sets.
	\label{def:sets}
	\begin{itemize*}
		\vspace{-0.5em}
		\item $\mathcal{D}^R$ is a multiset of node pairs connected by score links for the supervised term (Equation~\eqref{loss1}).
		\item $\mathcal{D}^+$ is a multiset of node pairs connected by similarity links, which is used in the positive unsupervised term (left term in Equation~\eqref{loss23}).
		\item $\mathcal{D}^-$ is a multiset of dissimilar node pairs
		for the negative unsupervised term (right term in Equation~\eqref{loss23}).
	\end{itemize*}
\end{mydef}

We extract node pairs from node sequences generated by the walks and assign the pairs into the entity pair sets according to the entities' relationships.
In a sliding window on each sequence, we appoint a center as $target$ and other members as $neighbors$.
We pair up the $target$ and each $neighbor$.
We assign the pairs into $\mathcal{D}^R$ if they are linked by score links, into $\mathcal{D}^+$ if the paired entities are similar and into $\mathcal{D}^-$ if the paired ones are dissimilar.
In the unweighted walk, we assign each pair into $\mathcal{D}^R$ for entities linked by score links and discard it otherwise.
In the positive walk, we assign each pair into $\mathcal{D}^R$ for entities of score links and into $\mathcal{D}^+$ otherwise.
In the negative walk, we assign each pair into $\mathcal{D}^R$ for entities linked with score links, into $\mathcal{D}^-$ for a user and an item, and into $\mathcal{D}^+$ for two users and two items.
Figure \ref{fig:ex2} shows an example of extracting node pairs.

We learn biases and vectors of all node $v \in \tilde{G}$ that minimize $L$ with the extracted node pairs in $\mathcal{D}^R, \mathcal{D}^+$ and $\mathcal{D}^-$ by gradient descent method with momentum \cite{qian1999momentum} for a faster learning.
The gradient update processes are stated in the supplementary material (Section 4.1 of \cite{supp}).


\paragraph{Step 3: Predict ratings.}
\label{sec:steppredict}
For a user $u$ and an item $i$, we predict rating of $i$ given by $u$ as $\hat{r}_{ui} = \mu + b_u + b_i + \mathbf{z}_u^T\mathbf{z}_i$.

\subsection{\method for Explainable Recommendation.}
\label{sec:explain}
For a user $u$, we recommend items $i$ whose predicted ratings $\hat{r}_{ui} = \mu + b_u + b_i + \mathbf{z}_u^T\mathbf{z}_i$ are the highest.
\method not only predicts rating scores,
but also explains why items are recommended to a user.
\method recommends items to \textit{a target user} $u$ because
(1) other users who are similar to $u$ like the items, and
(2) $u$ likes other items that are similar to the recommended items.
We reinforce the above two explanations by further examining the \textit{supplementary entities} which are defined as follows.
\vspace{-0.5em}
\begin{mydef} Supplementary entity.\\
	Supplementary entities are entities used in explanation of recommendation such as other users similar to a target user, and items similar to the recommended items.
\end{mydef}
\vspace{-0.5em}

We explain why the \textit{supplementary entities} are actually similar to $u$ and the recommended items for more understandable explanations.
The further explanation is stated as follows.
\vspace{-0.3em}
\begin{mydef} Meta-explainability of \method.\\
	\method provides meta-explanation (explanation for explanation): it explains why the supplementary entities are similar to target user or recommended items.
\end{mydef}
\vspace{-0.3em}

We use $\mathcal{D}^R$ and  $\mathcal{D}^+$ in our explanation, because we use rating information to denote like (high rating) or dislike (low rating), and similar relationships of entities.
We do not use $\mathcal{D}^-$ in explanation, because multi-step dissimilar relationships are not used.
The only dissimilar relationship we use is from low observed ratings in $\mathcal{D}^R$ such as common unlikable items in Figure \ref{fig:subreason1}.

The first reason of our recommendations is that other similar users prefer the recommended items.
The similar users are defined as ones with the highest similarities to the target user among all candidates who rate at least one of the recommended items.
A similarity of distinct entities $v$ and $w$ is defined in Equation \eqref{eq:sim},
where $\#(v, w)$ denotes the number of  times the pair $(v, w)$ appears in $\mathcal{D}^+$, and $\#v = ( \sum_{(v, w) \in \mathcal{D}^+}1 ) + ( \sum_{(w, v) \in \mathcal{D}^+}1 )$.
\vspace{-0.7em}
\begin{equation}
\small
\label{eq:sim}
sim(v, w) =
\frac
{\#(v, w)}
{\#v \#w}
\end{equation}
$\mathcal{D}^+$ is a set of similar entities defined in Definition \ref{def:sets}.
$\#v \#w$ serves as a normalization term.

The next reason of our recommendations is that $u$ prefers other similar items.
The similar items are selected among all candidates that $u$ rated.
For each recommended item $i$, we find out the most similar items $j$ to the recommended item among the candidates with the similarity score $sim(i, j)$ defined in Equation \eqref{eq:sim}.

\method also provides meta-explanations.
We explain why the supplementary entities are similar to the target user or the recommended items.
The supplementary users are determined to be similar because the supplementary users and the target user have common friends, common favorite items, or common unlikable items.
The common favorite items are highly rated items by the users, and the common unlikable items are lowly rated ones in observed ratings.
The supplementary items are determined to be similar items because common users rate them favorably.

\section{Experiment}
\label{sec:experiment}
We present experimental results to answer the following questions.
\begin{itemize*}
	\item \textbf{Q1 (Explainability)}: How can \method explain its recommendation results to users? (Section \ref{sec:exp:explain})
	\item \textbf{Q2 (Accuracy)}: How accurately does \method predict ratings? (Section \ref{sec:accuracy})
	\item \textbf{Q3 (Learning details)}: How does accuracy of \method change during iterations? How do hyperparameters affect the accuracy? (Section \ref{sec:learning_process})
\end{itemize*}

\subsection{Experimental settings.}
\label{sec:expsetting}
\noindent\textit{\textbf{\\Machine.}}
All experiments are conducted with a single CPU Intel(R) Xeon(R) CPU E5-2640 v3 @ 2.60GHz with 32 GB memory.

\noindent\textit{\textbf{\\Datasets.}}
We use social rating network data that have both observed ratings and a social network.
We summarize datasets in Table~\ref{table:datasetstat}.
These datasets contain explicit ratings and directed or undirected edges between users.
We convert directed edges to undirected edges in our experiment.

\begin{itemize*}
	\item \textit{FilmTrust}\footnote{ \url{https://www.librec.net/datasets.html}.} is a movie sharing and rating website.
	This dataset is crawled by Guo et al. \cite{guo2013novel}.
	
	\item \textit{Epinions}\footnote{\url{http://www.trustlet.org/epinions.html}} (\url{epinions.com})
	is a who-trust-whom online social network of general consumer reviews.
	
	\item \textit{Flixster}\footnote{\url{https://www.librec.net/datasets.html}} (\url{flixster.com})
	is also a website where users share movie ratings.
	It is crawled by Jamali and Etser \cite{jamali2010matrix}.
	
\end{itemize*}

\noindent\textit{\textbf{Competitors.}}
We compare \method with explainable methods UCF and ICF.
We also compare \method with MF to show if the unsupervised terms in \method improve accuracy.
In addition, we choose TrustSVD as a baseline method because it outperforms other social recommender systems.
\begin{itemize*}
	\item
		UCF (User-based collaborative filtering) predicts a rating of a user $u$ with $k$ most similar users of cosine the highest cosine similarity.
		A predicted rating is calculated as $\hat{r}_{ui} = \frac{1}{k}\sum_{v \in K_k}r_{vi}$, where $K_k$ is a set of the similar users.
		UCF explains its recommendation: it recommends items that are preferred by the similar users.

	\item
		ICF (Item-based collaborative filtering) predicts a rating of an item $i$ with $k$ most similar items in terms of cosine similarity with $i$.
		The rating is predicted as $\hat{r}_{ui} = \frac{1}{k}\sum_{j \in K_k}r_{uj}$, where $K_k$ is a set of the similar items.
		ICF explains its recommendations for a user $u$: it recommends items that are similar to $u$'s favorite items.

	\item MF (Matrix factorization with bias terms) is described in Section \ref{sec:mf}.
	
	\item TrustSVD \cite{guo2015trustsvd} is based on matrix factorization, and the state-of-the-art method among the ones that use both a rating matrix and a social network.
	TrustSVD outperforms other methods including
	PMF \cite{mnih2008probabilistic},
	RSTE \cite{ma2009learning},
	SoRec \cite{ma2008sorec},
	SoReg \cite{ma2011recommender},
	SocialMF \cite{jamali2010matrix},
	TrustMF \cite{yang2013social},
	and SVD++ \cite{koren2008factorization}.

\end{itemize*}

\noindent\textit{\textbf{Hyperparameters.}}
The hyperparameters of \method, MF, UCF, and ICF were determined experimentally, and they are reported in supplementary document.
We use hyperparameters for \method as follows.
In \textit{Filmtrust} dataset, $c$=5, $l$=30, $\alpha$=0.05, $\beta$=0.005, $d$=25, $s$=7, $\lambda_b$=0.1, $\lambda_z$=0.1, $\eta$=0.01, and $\gamma$=0.2.
In \textit{Epinions} dataset, $c$=6, $l$=50, $\alpha$=0.001, $\beta$=0.0007, $d$=25, $s$=7, $\lambda_b$=0.08, $\lambda_z$=1.3, $\eta$=0.003, and $\gamma$=0.6.
In \textit{Flixster} dataset, $c$=5, $l$=50, $\alpha$=0.001, $\beta$=0.001, $d$=25, $s$=7, $\lambda_b$=0.2, $\lambda_z$=0.2, $\eta$=0.005, and $\gamma$=0.6.
We use hyperparameters for TrustSVD as reported in \cite{guo2015trustsvd}.

\begin{table}[t]
	\centering
	\caption{Statistics of Datasets.}
	\renewcommand{\tabcolsep}{2mm}
	\small
	\begin{tabular}{l r r r}
		\toprule
		& \textit{FilmTrust}  	& \textit{Epinions}			 &\textit{Flixster}	\\
		\midrule
		\# of users        		  & 1,642						& 49,289				& 787,213 \\
		\# of items       	 	  & 2,071						& 139,738				& 48,794 \\
		\# of rating    	    & 35,494					& 664,824				& 8,196,077 \\
		\# of social edges  & 1,309					& 381,036				& 7,058,819 \\
		\bottomrule
	\end{tabular}
	\captionsetup{width=0.5\textwidth}
	\label{table:datasetstat}
\end{table}


\subsection{Explainability of \method.}
\label{sec:exp:explain}
We show an experiment on our explanation approach with \textit{Filmtrust} dataset.
\method explains the reason behind its recommendations in two ways with meta-explanations for each reason.
Items are recommended, because they are preferred by other similar users, and the target user likes other items that are similar to the recommended ones.
We generate meta-explanation about the reasons by providing further explanation on supplementary entities.
We explain why the \textit{supplementary} users and items are determined to be similar.

\begin{figure}[t]
	\centering
	\begin{subfigure}[b]{0.23\textwidth}
		\centering
		\includegraphics[width=0.9\textwidth]{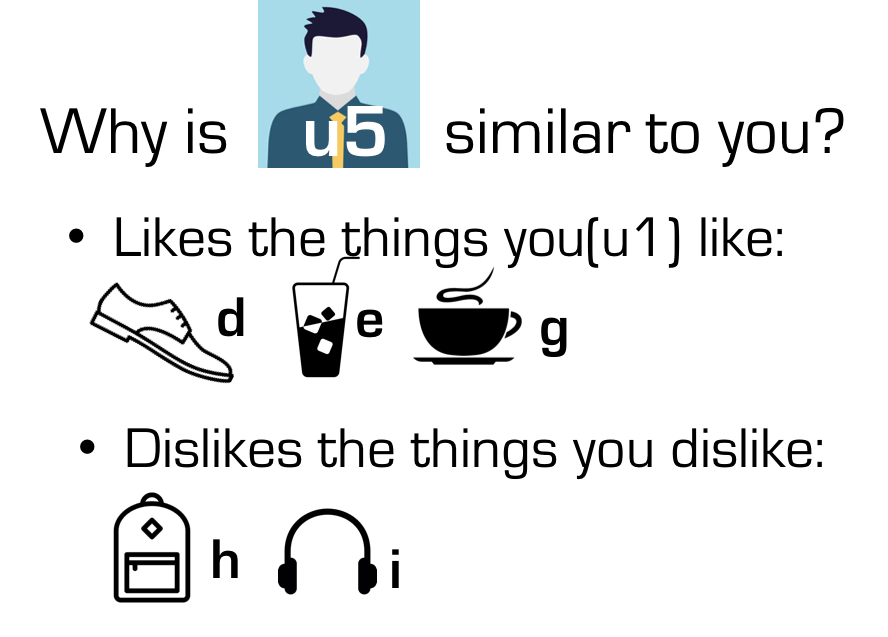}
		\caption{
			$u5$ is determined to be similar to the target user ($u1$), because they like or dislike common items.
		}
		\label{fig:subreason1}
	\end{subfigure}
	~
	\begin{subfigure}[b]{0.23\textwidth}
		\centering
		\includegraphics[width=0.9\textwidth]{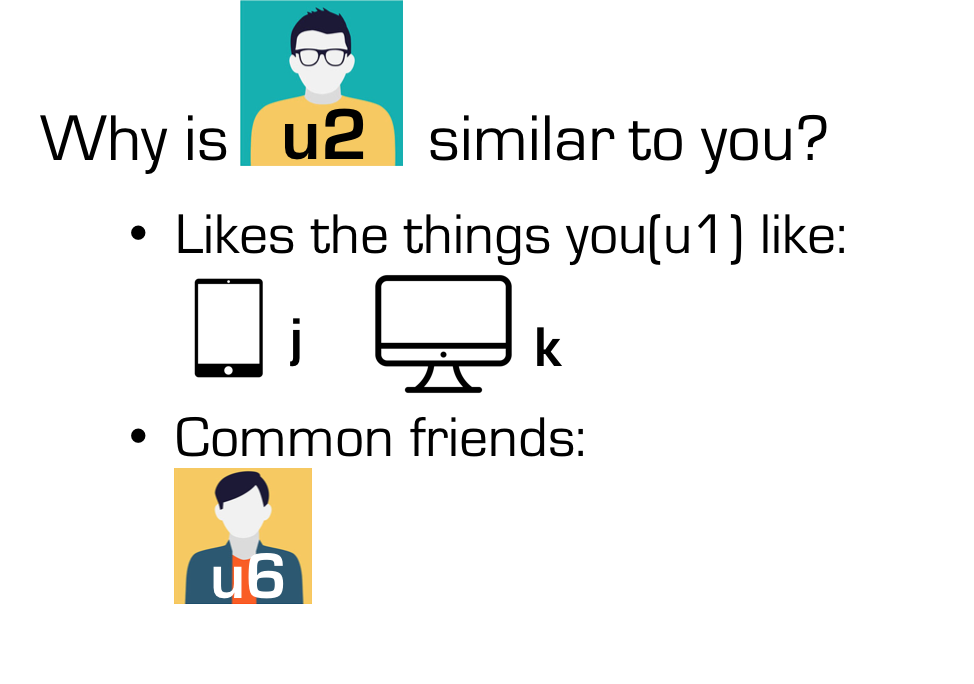}
		\caption{
			$u2$ is determined to be similar to $u1$, because they have common favorite items and friends.
		}
		\label{fig:subreason2}
	\end{subfigure}
	\begin{subfigure}[b]{0.5\textwidth}
		\centering
		\includegraphics[width=0.6\textwidth]{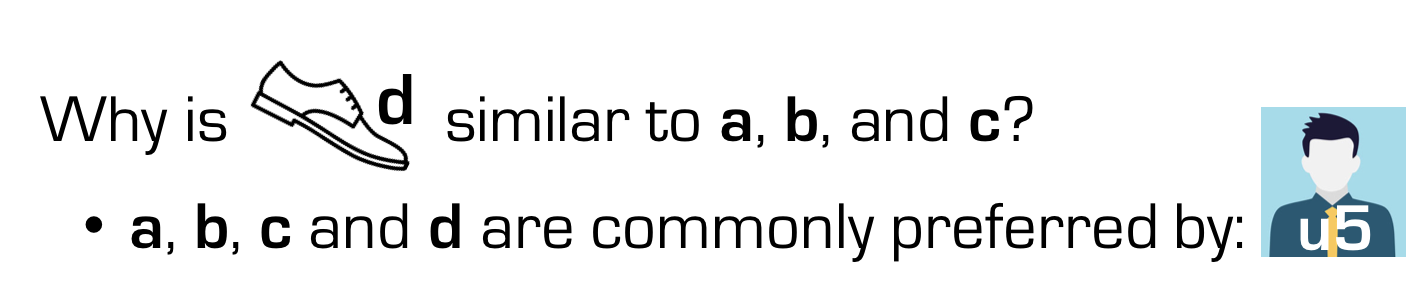}
		\caption{
			The supplementary item $d$ is similar to the recommended items since they are commonly preferred by other users.
		}
		\label{fig:subreason3}
	\end{subfigure}
	\caption{
		\method's meta-explanation to further explain why the \textit{supplementary entities} are determined to be similar to target user ($u1$) and recommended items ($a$, $b$, and $c$) in Figure \ref{fig:Explainability}.
	}
	\label{fig:meta-explainability}
\end{figure}

Figure \ref{fig:Explainability}\footnote{Ids of entities in \textit{Filmtrust} dataset: $u1$=28, $u2$=1332, $u3$=814, $u4$=319, $u5$=1038, $a$=1803, $b$=2820, $c$=3635, $d$=1859, $e$=1919, $f$=1948.} illustrates our explanation of recommendation.
Figure \ref{fig:reason1} illustrates the first reason in our recommendation that the items ($a$, $b$, and $c$) are recommended to the target user ($u1$): other similar users ($u2$, $u3$, $u4$, and $u5$) like them.
We describe the similar users more by presenting their similarity and friend relationships with the target user.
Figures \ref{fig:subreason1}\footnote{$g$=1700, $h$=1908, $i$=1921.} and \ref{fig:subreason2}\footnote{$u6$=508, $j$=1930, $k$=1933.} explain why the similar users are determined to be similar.
They like or dislike common items and have common friends.
Figure \ref{fig:reason2} explains the second reason why the items are recommended: the target user likes other similar items ($d$, $e$, and $f$).
We explicitly present similarity scores of \textit{supplementary items} with the recommended items.
Figure \ref{fig:subreason3} explains why the supplementary item ($d$) is determined to be similar to the recommended items.
It is preferred by common people including the target users' friends such as $u5$.

\vspace{-1em}
\subsection{Accuracy.}
\label{sec:accuracy}
We measure RMSE (Root Mean Square Error) and MAE (Mean Absolute Error) to evaluate accuracy of recommendation models.
Lower values indicate accurate predictions.
We separate training and test sets under 5-folded cross validation, ensuring that each observed rating is included in a test set only once.

Tables~\ref{table:RMSE} and~\ref{table:MAE} show RMSE and MAE of \method and competitors, respectively.
\method shows the lowest RMSE and MAE, outperforming competitors and achieving the state-of-the-art accuracy for recommendation with rating data and a social network.
TrustSVD shows the second best accuracy; however, it fails to explain its recommendation results, as we described in Section~\ref{sec:exp:explain}.
\method is much more accurate than MF, which indicates our extended unsupervised terms enhance accuracy.

\begin{table}[t]
	\centering
	\caption{\small RMSE of \method and other competitors.}
	\renewcommand{\tabcolsep}{1.0mm}
	\label{table:RMSE}
	\small
	\begin{tabular}{c c c c c c}
		\toprule
		\multirow{2}{*}{Dataset} & \textbf{\method}    & \multirow{2}{*}{ \textbf{UCF}} & \multirow{2}{*}{\textbf{ICF}} & \multirow{2}{*}{\textbf{MF}} & \multirow{2}{*}{\textbf{TrustSVD}}\\
		& \textbf{(proposed)} &                           &                     &                   &   \\ \midrule
		\textit{Filmtrust}	& \textbf{0.783} & 0.924	& 0.914 & 0.839 & 0.787\\
		\textit{Epinions}	& \textbf{1.041} & 1.200	& 1.257 & 1.135 & 1.043\\
		\textit{Flixster}	& \textbf{0.913} & 1.097	& 1.092 & 0.951 & 0.948\\ \bottomrule
	\end{tabular}
\end{table}

\begin{table}[t]
	\centering
	\caption{\small MAE of \method and other competitors.}
	\renewcommand{\tabcolsep}{1.0mm}
	\label{table:MAE}
	\small
	\begin{tabular}{c c c c c c}
		\toprule
		\multirow{2}{*}{Dataset} & \textbf{\method}    & \multirow{2}{*}{ \textbf{UCF}} & \multirow{2}{*}{\textbf{ICF}} & \multirow{2}{*}{\textbf{MF}} & \multirow{2}{*}{\textbf{TrustSVD}}\\
		& \textbf{(proposed)} &                           &                     &                   &   \\ \midrule
		\textit{Filmtrust}	& \textbf{0.598} &   0.727  & 0.713 & 0.658 & 0.607\\
		\textit{Epinions}	& \textbf{0.798} & 0.907 & 0.939 & 0.873 & 0.804\\
		\textit{Flixster}	& \textbf{0.711} & 0.854	& 0.871 & 0.739 & 0.726\\ \bottomrule
	\end{tabular}
\end{table}

\subsection{Learning Details.}
\label{sec:learning_process}
We present details of \method's learning process and hyperparameter sensitivity.

\begin{figure}[t]
	\centering
	\includegraphics[width=0.45\textwidth]{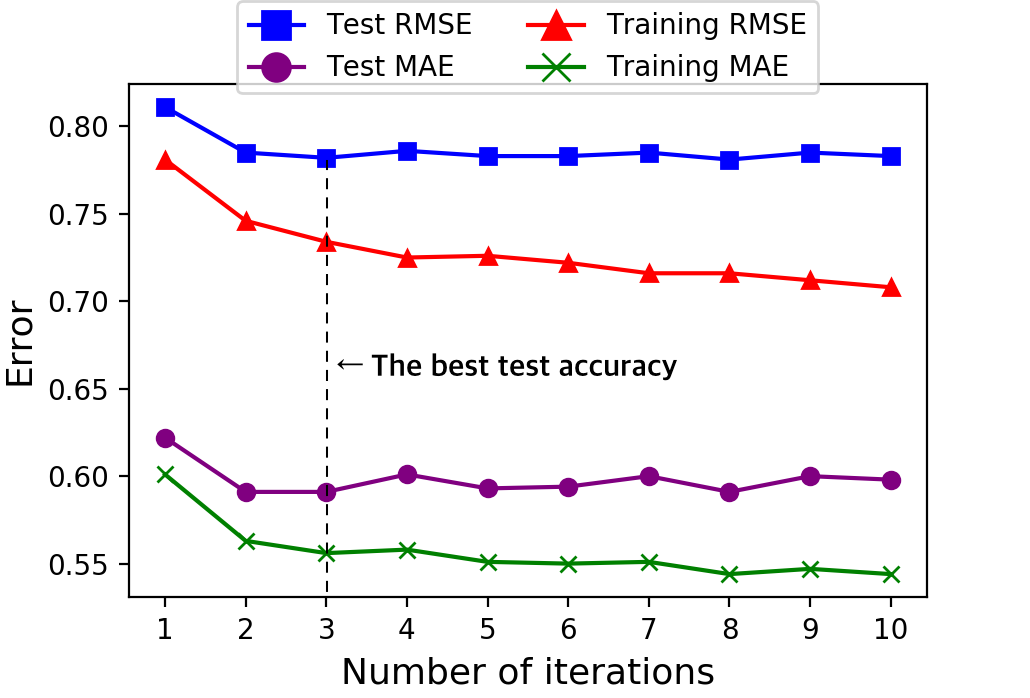}
	\caption{Changing accuracy of \method in training and test set. RMSE and MAE decrease initially and converge through the iterations.}
	\label{fig:learning_process}
\end{figure}

\begin{figure}[t]
	\centering
	\begin{subfigure}[b]{0.25\textwidth}
		\centering
		\includegraphics[width=1\textwidth]{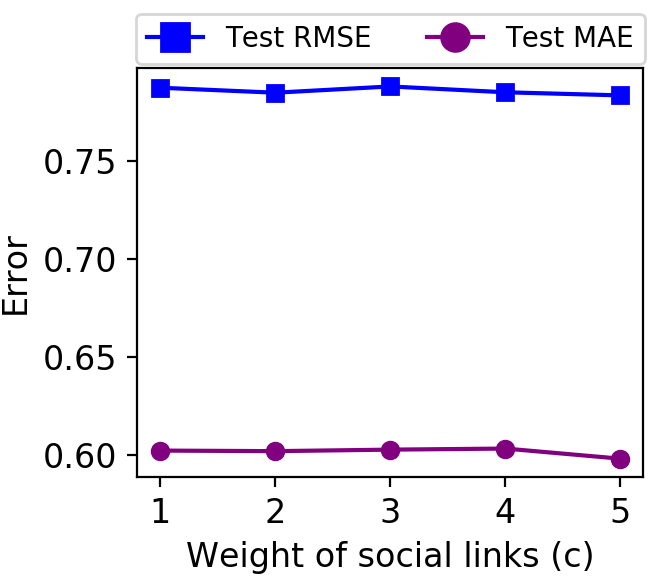}
		\caption{Sensitivity to $c$.}
		\label{fig:sens_c}
	\end{subfigure}
	~
	\begin{subfigure}[b]{0.22\textwidth}
		\centering
		\includegraphics[width=1.05\textwidth]{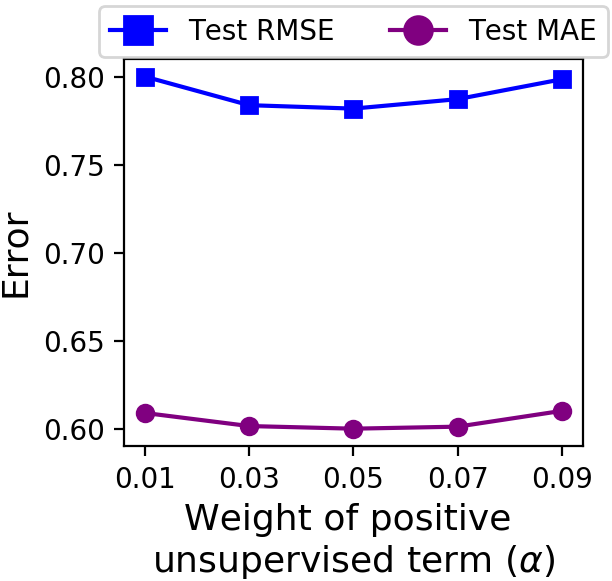}
		\caption{Sensitivity to $\alpha$.}
		\label{fig:sens_alpha}
	\end{subfigure}
	\begin{subfigure}[b]{0.5\textwidth}
		\centering
		\includegraphics[width=1.05\textwidth]{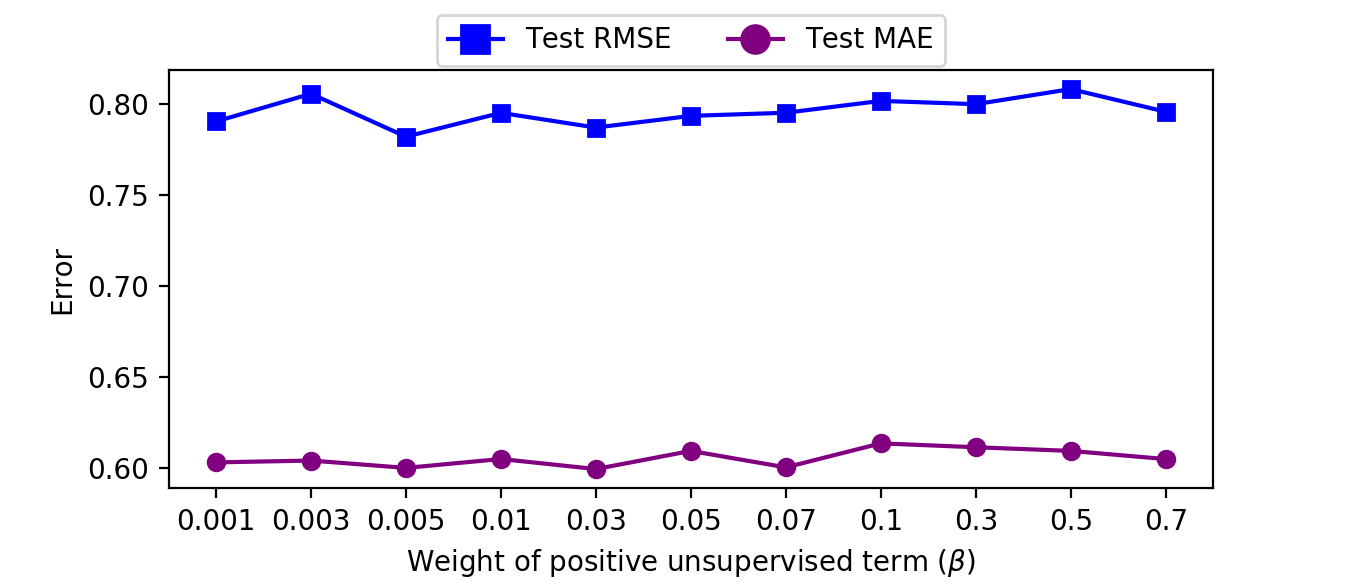}
		\caption{Sensitivity to $\beta$.}
		\label{fig:sens_beta}
	\end{subfigure}
	\caption{
		\method's hyperparameter sensitivity.
	}
	\label{fig:sensitivity}
\end{figure}

We first present how the accuracy changes during learning iterations: a iteration performs learning from positive walks, negative walks, and unweighted walks (lines 9-11 in Algorithm 2 of \cite{supp}).
Figure \ref{fig:learning_process} shows the learning process of \method on \textit{FilmTrust} dataset.
We note that the minimum test error (denoted by the dashed line) is achieved after only three iterations, thanks to the momentum term in the gradient descent
(Section 4.1 of \cite{supp}).
Learning with the momentum is 1.45$\times$ faster and requires 2.5$\times$ less iterations than learning without the momentum in our experiment.
We assess our model's sensitivity to the learning weight $\alpha$ of positive unsupervised term, the learning weight $\beta$ of negative unsupervised term, and the weight $c$ of social links; sensitivity for other hyperparameters is reported in the supplementary document (Section 6.3 of \cite{supp}).
Figure \ref{fig:sensitivity} shows the \method's hyperparameter sensitivity on \textit{FilmTrust} dataset.
\method is less sensitive to $c$.
\method is sensitive to $\alpha$ and $\beta$, but not much.

\section{Related Works}
\label{sec:related}
We review related works along the following highly related aspects: recommender systems with auxiliary information, explainable recommendation, and network embedding.

\subsection{Recommendation with Additional Data.}
Many studies have proved that using auxiliary data in recommendation improves accuracy.
Various additional data alleviate rating sparsity problem.
Recent studies employ auxiliary information for better performance.

\textit{Social Recommendation.}
Many recommender systems use social networks which contain rich information on influences among users.
Studies  \cite{ma2008sorec, huang2010social, yang2012circle, jiang2012social} prove that a social network enhances rating prediction.
\cite{Zhao2014LeveragingSC} trains ranking-based model with a social network.
Chaney et al. \cite{chaney2015probabilistic} develop a probabilistic model that integrates social network with traditional matrix factorization method.

\textit{Heterogeneous Information.}
Other information is used to improve recommendation performance.
Li et al. \cite{li2015point} use both geographical and social information.
Another example shows that attributes of entities (e.g. genre) are used in recommending personalized entities \cite{yu2014personalized}.
\cite{Zhang2016collaborative} divides heterogeneous data into three categories and extracts features from the categories.
Jeon et al.~\cite{DBLP:conf/icde/JeonJSK16} and Choi et al.~\cite{DBLP:journals/corr/abs-1708-08640} exploit coupled tensor factorization to use additional heterogeneous information in tensor based recommendation.

\vspace{-1em}
\subsection{Explainable Recommendation.}
Explainable recommender system explains reasons behind its recommendations.
A reasonable explanation is beneficial:
explainable predictions increase users' trust \cite{ribeiro2016should},
and induce the users to inform the system of its wrong predictions
\cite{zhang2014explicit}.
We study explainable recommendation models using users' reviews or social network.

\textit{Review-based explanation of recommendation.}
User reviews are useful in explaining recommendation because the reviews have keywords of users' preferences.
Many models such as SCAR \cite{ren2017social}, TriRank \cite{he2015trirank}, and EFM \cite{zhang2014explicit} exploit user reviews.
SCAR explains its recommendations by listing keywords of an item's concept with topic analysis, sentiment prediction, and viewpoint regression.
TriRank analyzes user-item-aspect tripartite graph, and explains its results using important phrases in the reviews to describe items.
EFM extracts feature words of products and users' opinions from reviews.
It suggests a set of feature words that users mainly focus on and that describe recommended products.

\textit{Friend-based explanation of recommendation.}
Social network is also useful to explain recommendations.
Recommendation models such as SPF \cite{chaney2015probabilistic} and LBSN \cite{kefalas2013new} explain reasons of its results with social links.
SPF recommends items to a user because the items are favorites of the users' friends.
LBSN recommends locations to a user at a certain time because the locations are visited by the users' friends and it is likely that the friends will visit the place again soon.

\vspace{-1em}
\subsection{Network Embedding.}
Network embedding has been studied extensively and has become a popular way to represent graph data.
Recently proposed network embedding methods \cite{perozzi2014deepwalk,tian2014learning,tang2015line,cao2015grarep,grover2016node2vec,cao2016deep} try to encode proximity structure.
Tian et al. \cite{tian2014learning} apply autoencoder to reconstruct neighborhood structure of each node.
Perozzi et al. \cite{perozzi2014deepwalk} propose DeepWalk model which applies language model on sequences of nodes generated by truncated random walks on a graph.
Tang et al. \cite{tang2015line} directly model first-order and second-order proximities and optimize them.
Grover and Leskovec \cite{grover2016node2vec} extend DeepWalk by adding parameters in the truncated random walk to customize the random walk on different data.

%



\vspace{-1em}
\section{Conclusion}
\label{sec:conclusion}
We propose \method, a novel explainable and accurate recommendation model that exploits both rating data and a social network.
\method constructs a unified graph containing users and items where the weights reflect the degree of association among users and items.
\method uses network embedding on the unified graph to extract latent features of users and items,
and predicts ratings with the embedded features.
\method provides the best explainability and accuracy for recommendation.
Future works include extending the method for distributed systems for scalable learning.

\bibliographystyle{siamplain}
\bibliography{sigproc}

\end{document}